\documentclass[sn-mathphys,iicol,Numbered]{sn-jnl}
\usepackage{graphicx}%
\usepackage{multirow}%
\usepackage{amsmath,amssymb,amsfonts}%
\usepackage{amsthm}%
\usepackage{mathrsfs}%
\usepackage[title]{appendix}%
\usepackage{xcolor}%
\usepackage{textcomp}%
\usepackage{manyfoot}%
\usepackage{booktabs}%
\usepackage{algorithmicx}%
\usepackage[ruled,vlined]{algorithm2e}
\usepackage{algpseudocode}%
\usepackage{listings}%
\usepackage{algpseudocode}
\usepackage{textcomp}
\usepackage{multirow}
\usepackage{lmodern}
\usepackage{bookmark}
 
\usepackage[normalem]{ulem}  

\renewcommand{\sout}{\bgroup \color{red} \ULdepth=-.5ex \ULset}

\usepackage{tikz,xcolor,hyperref}
\definecolor{lime}{HTML}{A6CE39}
\DeclareRobustCommand{\orcidicon}{
	\begin{tikzpicture}
	\draw[lime, fill=lime] (0,0) 
	circle [radius=0.16] 
	node[white] {{\fontfamily{qag}\selectfont \tiny ID}};
	\draw[white, fill=white] (-0.0625,0.095) 
	circle [radius=0.007];
	\end{tikzpicture}
	\hspace{-2mm}
}

\foreach \x in {A, ..., Z}{%
	\expandafter\xdef\csname orcid\x\endcsname{\noexpand\href{https://orcid.org/\csname orcidauthor\x\endcsname}{\noexpand\orcidicon}}
}

\begin{document}

\title[Article Title]{Schr\"{o}dinger Generator for High-Dimensional  Integration and Sampling on Quantum Many-Body States}

\author[1,2]{\fnm{Lin-Jing} \sur{Jiang} \orcidA{}}  

\author[1,2]{\fnm{Fu} \sur{Ma}\orcidB{}}

\author[1,2]{\fnm{Pei} \sur{Li}\orcidC{}}  

\author*[1,2]{\fnm{Kai-Jia} \sur{Sun} \orcidD{}}\email{kjsun@fudan.edu.cn}

\author*[1,2]{\fnm{Guo-Liang}   \sur{Ma}  \orcidE{}}\email{glma@fudan.edu.cn} 
	
\author*[1,2,3]{\fnm{Yu-Gang} \sur{Ma}  \orcidF{}}\email{mayugang@fudan.edu.cn}


\affil[1]{\orgdiv{Key Laboratory of Nuclear Physics and Ion-beam Application (MOE), Institute of Modern Physics}, \orgname{Fudan University}, \orgaddress{\city{Shanghai}, \postcode{200433}, \country{China}}}
\affil[2]{\orgdiv{Shanghai Research Center for Theoretical Nuclear Physics}, \orgname{NSFC and Fudan University}, \orgaddress{\city{Shanghai}, \postcode{200438}, \country{China}}}

\affil[3]{\orgdiv{School of Physics}, \orgname{East China Normal University}, \orgaddress{\city{Shanghai}, \postcode{200241}, \country{China}}}

\abstract{Integration and sampling in high dimensions are among central challenges in modern science and technology, underlying applications ranging from quantum many-body physics to Bayesian inference and artificial intelligence. Although conventional Monte Carlo methods are formally scalable, their efficiency deteriorates rapidly in the presence of strong correlations or sharp features in high-dimensional configuration space. Here we introduce a new framework, termed the ``Schr\"{o}dinger Generator'', for integration and sampling based on the explicit optimization of coordinate transformations. The method decomposes the total Jacobian into two complementary components, an adaptive map that reduces estimator variance by learning marginal structures and a normalizing-flow transformation that captures non-factorizable correlations. A final resampling step ensures unbiased sampling despite imperfections in the learned transformation. We demonstrate stable and scalable performance beyond 600 dimensions, faithfully reproducing short-range nucleon-nucleon correlations in finite nuclei. The framework provides a physically transparent approach to high-dimensional integration and sampling of complex quantum systems.

}




\maketitle


\begin{figure*}[!t]
\centering 
\includegraphics[width=\linewidth]{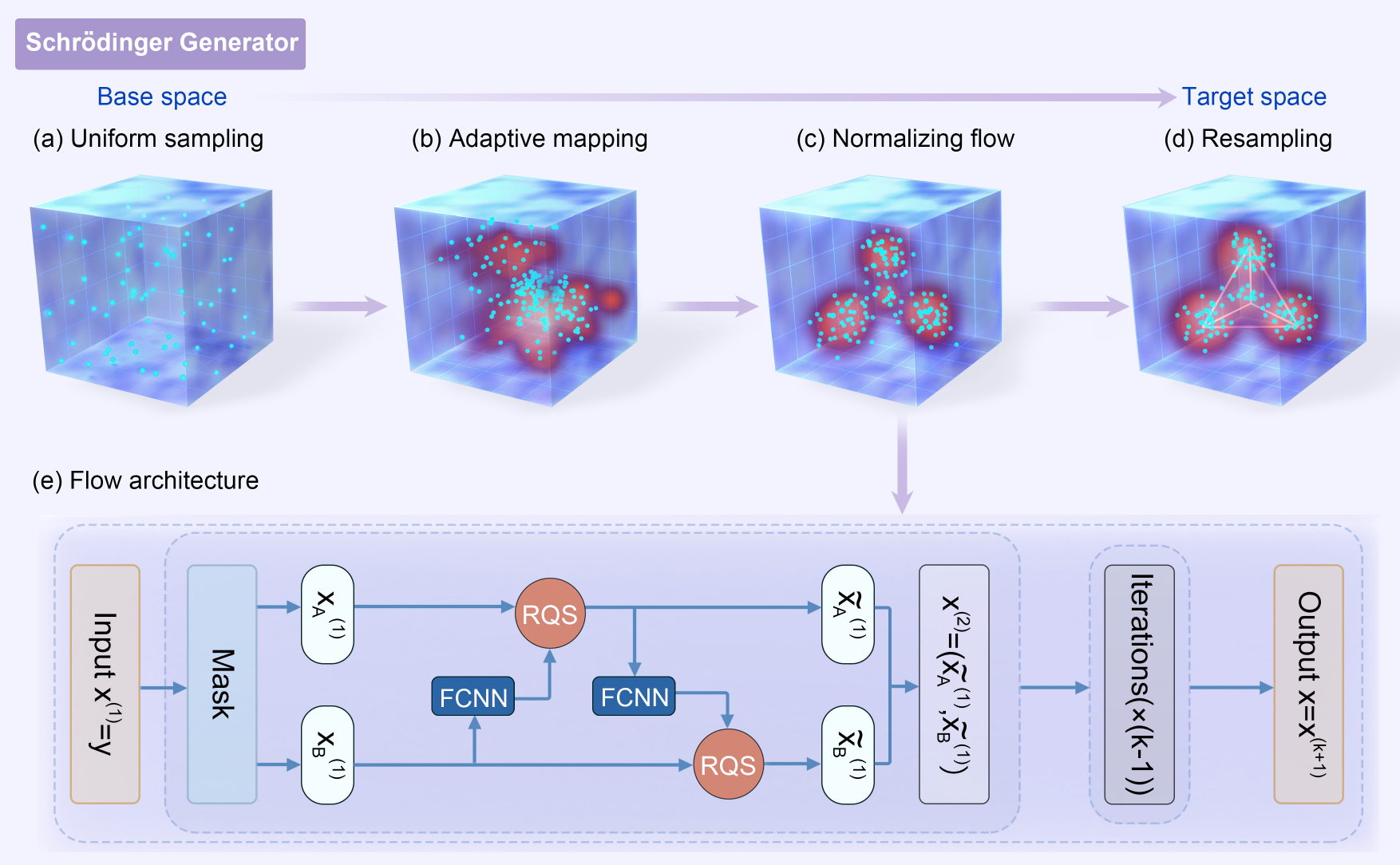} 
\caption{
\textbf{Schematic of the Schr\"odinger Generator for high-dimensional integration and sampling.} 
The framework combines an adaptive coordinate transformation, which reduces variance in each dimension, with a normalizing-flow transformation, which learns high-dimensional correlations, enabling efficient integration and low-variance sampling from complex target distributions. 
Upper panels show evolution of density distribution at successive stages: uniform initial sampling (a), adaptive importance mapping (b), flow-based transformation (c), and final resampling (d). Lower panels show architecture of the normalizing flow transformation using rational quadratic spline (RQS) with fully connected neural network (FCNN) (e).
}
\label{pic:SG} 
\end{figure*}
High-dimensional integration and sampling are of fundamental importance to first-principles calculations across physics, chemistry, computer science, and etc~\cite{kalos2008monte}. In quantum many-body physics, observables are often expressed as integrals over exponentially large configuration or phase spaces, while accurate sampling of strongly correlated distributions is essential for computing expectation values and fluctuations~\cite{Joseph:2019zer}. In high-energy physics, the increasingly precise experimental measurements from current and future runs of facilities such as the Large Hadron Collider (LHC) demand a corresponding level of accuracy in theoretical simulations~\cite{HSFPhysicsEventGeneratorWG:2020gxw,ATLAS:2021yza}. In nuclear physics, sampling initial nucleon distributions from the nuclear many-body wavefunction becomes indispensable for imaging nuclear shapes with high-energy nuclear collisions~\cite{STAR:2024wgy}.

Deterministic integration methods suffer from the curse of dimensionality, with computational cost growing exponentially in dimension, rendering them impractical for many-body problems. Consequently, Monte Carlo (MC) techniques remain the standard approach~\cite{Rosenbluth1955,MacGillivray1982,szirmaykalos2008monte,Rogers_2006,jaeckel2002monte,Webber:1986mc,Buckley:2011ms}. However, their efficiency deteriorates in high dimensions due to large variance, especially for sharply peaked or strongly correlated distributions. Classical variance-reduction techniques, including importance sampling and adaptive algorithms such as \textsc{VEGAS}~\cite{Lepage:1977sw,Lepage:2020tgj}, \textsc{FOAM}~\cite{Jadach:2002kn}, and related frameworks, become increasingly ineffective as correlations grow more intricate. In particular, separable and axis-aligned adaptations can generate “phantom” peaks whose number grows rapidly in higher dimensions, thereby degrading sampling efficiency~\cite{Lepage:2020tgj}.

Machine learning offers promising alternatives. Generative models such as GANs, VAEs, and especially normalizing flows (NFs)~\cite{Tabak:2010vsy,Tabak:2013cnz,Dinh:2014mzt,rezende2015variational} can learn complex high-dimensional distributions~\cite{Bendavid:2017zhk,Klimek:2018mza,Otten:2019hhl,Butter:2019cae}. In particular, normalizing flows provide invertible transformations with tractable Jacobians and have been successfully integrated with MC methods~\cite{song2018anicemcadversarialtrainingmcmc,Levy:2017bji,Gao:2020zvv,Bothmann:2020ywa,Winterhalder:2021ngy,Heimel:2022wyj,Butter:2022rso}. Nevertheless, directly learning the full target density can be problematic, as issues such as mode collapse, incomplete support coverage, and topological mismatch may arise, particularly for distributions featuring sharp nodes or constrained manifolds~\cite{Kobyzev_2021,cornish2021relaxingbijectivityconstraintscontinuously}.

To address limitations in the above-mentioned methods, we introduce the Schrödinger Generator (SG), a hybrid framework that explicitly optimizes the Jacobian of a learned coordinate transformation. It combines an adaptive mapping that reduces variance along individual dimensions and learns marginal distributions, with a flow-based transformation that captures correlations across different dimensions. This combined transformation greatly reduces variance in the estimation of integration of the target distribution. A final resampling step guarantees unbiased sampling even when the learned transformation is imperfect. The overall workflow of SG and the architecture of normalizing flow are illustrated in Fig.~\ref{pic:SG}.

We benchmark the Schrödinger Generator on integration and sampling tasks for  quantum many-body states. The method efficiently merges the phantom peaks generated by the adaptive importance sampling into the true modes of the target distribution. Because all true modes are encoded in the product of marginal distributions and are fed into the normalizing flows, the Schrödinger Generator does not exhibit mode collapse. Compared with \textsc{VEGAS}, it achieves significantly higher accuracy and lower Kullback–Leibler divergence. We further apply the method to nuclear quantum states up to dimensions of $D=624$, where the generated samples faithfully reproduce short-range nucleon correlations. The Schrödinger Generator thus establishes a scalable and physically interpretable framework for high-dimensional stochastic integration and sampling, paving the way for numerical simulations of strongly correlated quantum systems across the nuclear chart and beyond.

\bmheadA{\Large Results} 
\bmhead{Strategy of the Schr\"{o}dinger Generator}
We start by considering the $D-$dimensional integral
\begin{equation}
I=\int \mathrm{d}^D{\bf x} f({\bf x}),~ {\bf x}\in \mathbb{R}^D
\end{equation}
and introduce an invertible mapping ${\bf x}={\bf x}({\bf z})$ with a Jacobian
$J_{\rm SG}({\bf z})=\left|\partial{\bf x}/\partial{\bf z}\right|$. Constructing a tractable, invertible, and expressive $J_{\rm SG}$ in high dimensions is a major challenge. For this purpose, we decompose the full transformation as successive transformations 
\begin{equation}
\mathbf{z}\xrightarrow{ \mathcal{T}_A }\mathbf{y}
\xrightarrow{ \mathcal{T}_F }\mathbf{x},
\end{equation}
where $\mathcal{T}_A$ implements adaptive variance reduction in each dimension and $\mathcal{T}_F$ further learns correlations using normalizing flows. This separation stabilizes optimization and assigns complementary roles to the two components.
For an arbitrary base density $p_0({\bf z})$, this gives
\begin{equation}
I=\int \mathrm{d}^D{\bf z}  p_0({\bf z})
\frac{f({\bf x} \circ {\bf y}({\bf z})) J_{\rm SG}({\bf z})}{p_0({\bf z})}
\end{equation}
with a factorized Jacobian
$J_{\rm SG}=J_A({\bf z})J_F({\bf y})$ and ${\bf x} \circ {\bf y}({\bf z})={\bf x}({\bf z})\equiv \mathcal{T}_F(\mathcal{T}_A({\bf z}))$.
Drawing $M$ samples ${\bf z}_i\sim p_0({\bf z})$, the Monte Carlo estimator of $I$ becomes
\begin{equation}
I_{\rm SG}
=\frac{1}{M}\sum_{i=1}^M
\frac{f({\bf x} \circ {\bf y}({\bf z}_i)) J_{SG}({\bf z}_i)}{p_0({\bf z}_i)}.
\end{equation}

To determine the first Jacobian $J_A$, we set the transformation $ \mathcal{T}_F$ to the identity, such that $J_F = 1$. In this case, $J_A$ is optimized according to the \textsc{VEGAS} algorithm by minimizing the variance of the Monte Carlo estimator, 
\begin{equation}
\sigma_I^2=\frac{1}{M}\left [\int \mathrm{d}^D{\bf z} 
p_0^{-1}({\bf z})J_{A}^2({\bf z})  f^2({\bf x}({\bf z}))
-I^2 \right ].
\label{eq:sigI}
\end{equation}
where $M$ is the number of independent Monte Carlo samples used to estimate the integral. Provided the variance is finite, the standard error scales as $M^{-1/2}$ which has no explicit dependence on the dimensionality $D$ and is thus advantageous compared to deterministic methods. However, it becomes inefficient when the integrand is sharply localized or exhibits strong correlations. Minimizing $\sigma_I^2$ with respect to the Jacobian yields, in one dimension and under normalization $J_A(z) \propto 1/{|f(x(z))|}$.
This result indicates that regions where the integrand is large are effectively compressed, while regions where it is small are stretched, thereby improving sampling efficiency.

\begin{figure*}[!t]
\centering 
\includegraphics[width=\linewidth]{ 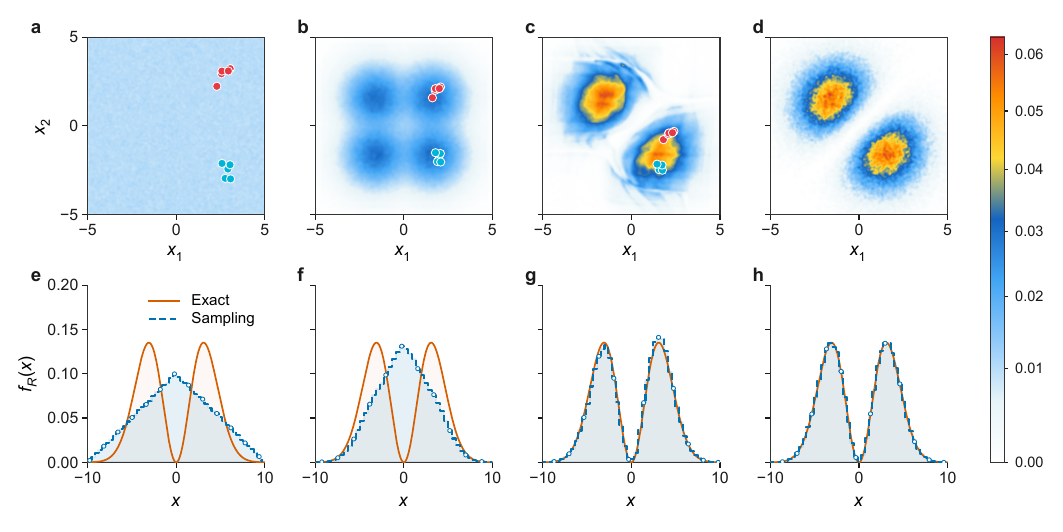} 
\caption{
\textbf{Evolution of the sampled density for the two-dimensional fermionic distribution.}
Density distributions at successive stages of the Schr\"{o}dinger generator: uniform initial sampling (a), adaptive importance mapping (b), normalizing-flow transformation (c), and final resampling (d). The red and blue solid circles in (a)-(c) denote representative samples. Distribution of relative distance $x=x_1-x_2$ obtained analytically (red solid lines) and from numerical samples (blue dashed lines) are presented in the corresponding lower panels. The SG progressively captures the bimodal structure and the suppressed diagonal region of the target distribution. 
}
\label{pic:2D} 
\end{figure*}

In practice, the optimization in Eq.~(\ref{eq:sigI}) is implemented by iteratively adjusting the bin widths of a $D$-dimensional grid along each coordinate. In this way, the learned transformation captures only the marginal structure of the target distribution.

After optimizing $J_A$, the second Jacobian $J_F$ is determined by minimizing the loss function e.g., Kullback-Leibler (KL) divergence~\cite{Kullback:1951zyt} between the output of normalizing flow and the target distribution,
\begin{eqnarray}
L_\text{SG} &=& D_{KL}(p_0({\bf z})~||~f({{\bf x}({\bf z}))J_{\rm SG}({\bf z})}) \notag \\ 
&=& \mathbb{E}_{\mathbf{z}\sim p_0}
 [-\log f(\mathbf{x}(\mathbf{z}))-\log J_F  \notag \\
 &&~~ ~~~~~~-\log J_A+\log p_0(\mathbf{z})].
\end{eqnarray}

To guarantee unbiased sampling when the learned transformation deviates from optimality, the Schr\"{o}dinger Generator applies a resampling step. Output samples $\mathbf{x}$ are assigned weights
\begin{equation}
w_{\rm SG}(\mathbf{x}) \propto
\frac{f({\bf x} \circ {\bf y}(\mathbf{z}))\,J_{\rm SG}(\mathbf{z})}{p_0(\mathbf{z})},
\end{equation}
and are resampled accordingly to match the target density. The Schr\"{o}dinger Generator thus serves as unbiased integrator and sampler, simultaneously. For details of the algorithm, see Methods sections.

\bmhead{Sampling quantum states in two dimensions}
For benchmark, we consider the $N$-body density of an antisymmetrized fermionic wave function $\Psi(\mathbf{x}_1,\ldots,\mathbf{x}_N)$ constructed as a Slater determinant given by
\begin{equation}
\Psi
= \frac{1}{\sqrt{\mathcal{N}}}
\det \left[
\begin{matrix}
\phi_1(\mathbf{x}_1) & \phi_1(\mathbf{x}_2) & \cdots & \phi_1(\mathbf{x}_N) \\
\phi_2(\mathbf{x}_1) & \phi_2(\mathbf{x}_2) & \cdots & \phi_2(\mathbf{x}_N) \\
\vdots & \vdots & \ddots & \vdots \\
\phi_N(\mathbf{x}_1) & \phi_N(\mathbf{x}_2) & \cdots & \phi_N(\mathbf{x}_N)
\end{matrix}
\right]
\end{equation}
with a normalization factor $\mathcal{N}$ and the single-particle wave function given by
$ \phi_i(x_j)
= \left( \pi b^2  \right)^{-3/4}
\exp\!\left[ -\frac{1}{2b^2} (\mathbf{x}_j - \mathbf{R}_i)^2 \right]$. 
Here $\mathbf{R}_i$ denote the Gaussian centers and $b$ is the harmonic-oscillator width. The $N$-body density distribution is then given by $f_N(\mathbf{x}_1,\ldots,\mathbf{x}_N) = |\Psi(\mathbf{x}_1,\ldots,\mathbf{x}_N)|^2$.
This construction generates a strongly correlated density with nodal surfaces imposed by fermionic antisymmetry, posing a stringent challenge for generative sampling in high dimensions.

We first benchmark the Schrödinger Generator in a two-dimensional case ($D=2$), where the two fermions live in one-dimensional space. The joint probability density $f_2\left(x_1, x_2\right)$ is given as
\begin{eqnarray}
f_2&=&\frac{1}{2\pi b^2\left(1-e^{-\frac{\left(R_1-R_2\right)^2}{2b^2}}\right)}\big(e^{-\frac{\left(x_1-R_1\right)^2}{b^2}-\frac{\left(x_2-R_2\right)^2}{b^2}}  \notag \\
&+&e^{-\frac{\left(x_2-R_1\right)^2}{b^2}-\frac{\left(x_1-R_2\right)^2}{b^2}}  \\
&-&2e^{-\frac{\left(x_1-R_1\right)^2+\left(x_2-R_1\right)^2}{2b^2}-\frac{\left(x_2-R_2\right)^2+\left(x_1-R_2\right)^2}{2b^2}}\big).\notag
\end{eqnarray}

The values of parameters are set as $b=2$ and $R_1=1$, $R_2=-1$ with arbitrary units. This distribution represents a superposition of two correlated Gaussian components with a subtraction term that induces a pronounced suppression along the diagonal, providing a good test for generative sampling. The distribution of relative coordinate $x=x_1-x_2$ follows
\begin{eqnarray}
f_R(x)&=&\frac{2e^{-\frac{ R^2+ x^2}{2b^2}}\mathrm{sinh}^2(\frac{R x}{2b^2})}{\sqrt{2\pi b^2} (1-e^{-\frac{ R^2}{2b^2 }})},
\end{eqnarray}
with $R=R_1-R_2$.

\begin{figure*}[!t]
\centering 
\includegraphics[width=\linewidth]{ 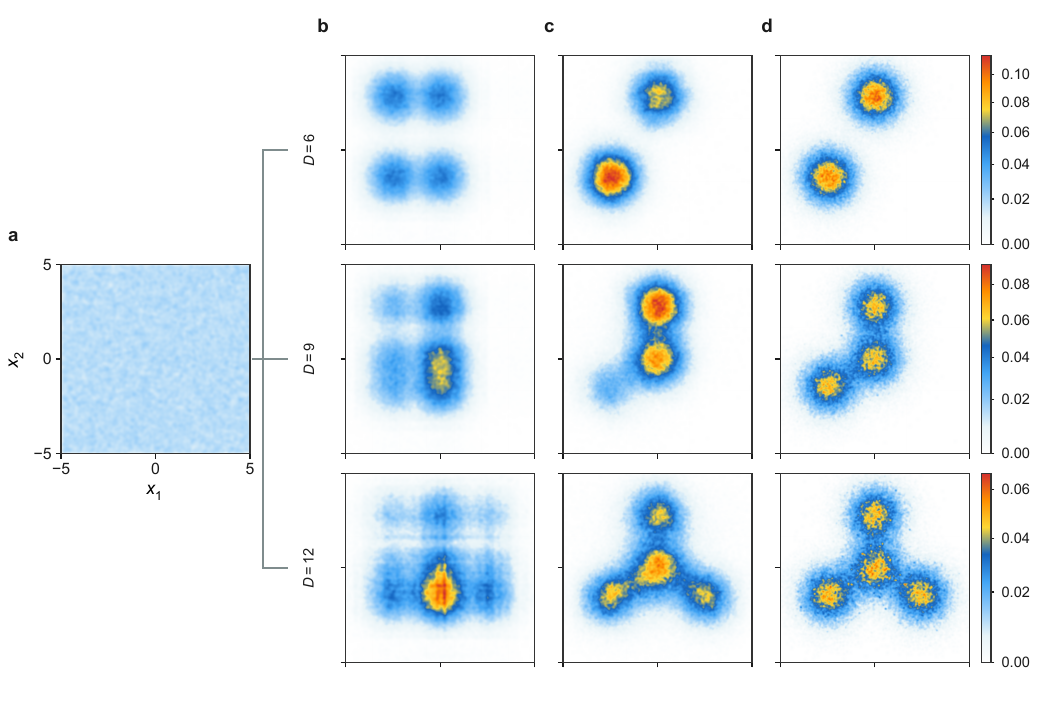}   
\caption{
\textbf{Learning fermionic densities with the Schr\"odinger Generator.}
Evolution of the sampled density at successive stages of the SG pipeline: uniform initialization (a), adaptive importance mapping (b), normalizing-flow transformation (c), and final resampling (d). Results are shown for dimensions $D=6$ (top), $D=9$ (middle) and $D=12$ (bottom).
}
\label{pic:multiD} 
\end{figure*}

Figure~\ref{pic:2D} illustrates the evolution of the sampled density during the different stages of the SG pipeline. Starting from an initially uniform distribution (Fig.~\ref{pic:2D} (a)), the adaptive coordinate transformation already captures the coarse structure of the target (Fig.~\ref{pic:2D} (b)), but leads to two additional phantom peaks. After the normalizing-flow transformation (Fig.~\ref{pic:2D} (c)), the generator reproduces both the bimodal structure and the suppressed region along $x_1\simeq x_2$. The final resampling step (Fig.~\ref{pic:2D} (d)) further sharpens the features and yields a distribution matching to the exact target distribution.

\begin{figure}[!t]
\centering 
\includegraphics[width=\linewidth]{ 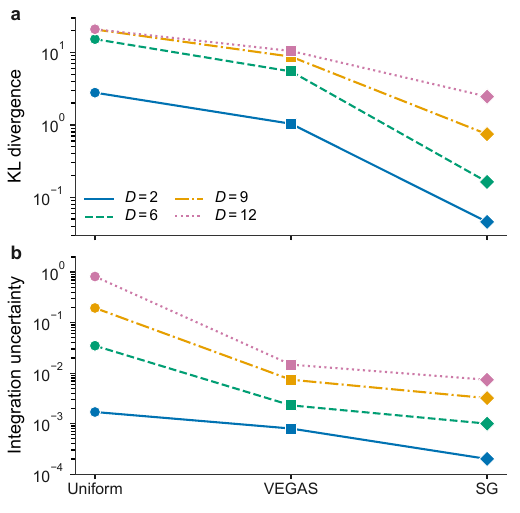} 
\caption{
\textbf{Comparison of sampling accuracy and integration efficiency.}
Upper panel: Kullback--Leibler divergence between the target distribution and samples obtained with uniform Monte Carlo, VEGAS sampling and the Schr\"odinger Generator for $D=2,\,6,\,9,\,12$. Lower panel: corresponding uncertainties of the integral with $10^6$ samples. 
}
\label{pic:compare} 
\end{figure}

The red and blue solid circles in Fig.~\ref{pic:2D} (a) are mapped onto the two peaks in panel (b) with red circles sitting on one phantom peak. After the mapping of normalizing flow, the two peaks merge together and both red and blue circles are mapped onto one real peak as shown in panel (c). In this way, the Schr\"{o}dinger Generator learns the non-linear correlations and eliminate the phantom peaks in conventional VEGAS sampling. Since all important modes already exist during the first-stage adaptive sampling, the successive transformation of normalizing flow learns to merge certain modes, and avoids mode collapse. 

Lower panels (e-h) of Fig.~\ref{pic:2D} depicts a quantitative comparison, where the distribution of relative distance $x=x_1-x_2$ obtained from output samples are confronted with the analytical results. In Fig.~\ref{pic:2D} (h), the numerical data (dashed lines) are in excellent agreement with the analytical curves (solid lines) over the full range, including the tails and the interference-induced dip. No systematic deviation is observed within statistical uncertainties, indicating that SG faithfully reproduces not only low-order moments but also the detailed shape of highly non-Gaussian structures.

\bmhead{Sampling quantum states in \texorpdfstring{$D=6,~9,~12$}{D=6, 9, 12} dimensions}
We next turn to higher-dimensional cases, fermions are now live in 3-dimensional space, where their center positions are given by $
\mathbf{R}_1 = \left( 0,\; 0,\; l \right),~ 
\mathbf{R}_2 = \left( -\frac{\sqrt{6}}{3}l,-\frac{\sqrt{2}}{3}l, -\frac{l}{3} \right), ~
\mathbf{R}_3 = \left( 0,\; \frac{2\sqrt{2}}{3}l,\; -\frac{l}{3} \right), ~
\mathbf{R}_4 = \left( \frac{\sqrt{6}}{3}l,\; -\frac{\sqrt{2}}{3}l,  -\frac{l}{3} \right)$.
The width and size parameters are taken as $b = 1.3$ and $l=3$, respectively. The normalized factors for $N=2$, 3, and 4 are given by $\mathcal{N}=2 \left(1- e^{-\frac{4l^2}{3 b^2}}   \right)$, $\mathcal{N}=6 \left(1- 3 e^{-\frac{4l^2}{3 b^2}} + 2 e^{-\frac{2l^2}{b^2}} \right)$, and $\mathcal{N}=24 \left(1- 3 e^{-\frac{8l^2}{3 b^2}} + 8 e^{-\frac{2l^2}{b^2}} - 6 e^{-\frac{4l^2}{3 b^2}}\right)$, respectively.

Figure~\ref{pic:multiD} shows the evolution of the sampled density for representative dimensions $D=6$, $9$, and $12$. Starting from uniform sampling, the adaptive mapping already concentrates probability mass in the physically relevant regions, while the subsequent normalizing-flow transformation captures the complex multimodal structure induced by fermionic antisymmetry. After the final resampling step, the SG samples reproduce the sharp peaks and suppressed nodal regions of the target density even at $D=12$, demonstrating that the method remains stable and expressive as dimensionality increases.

Comparing the second and third column panels of Fig.~\ref{pic:multiD}, one can see that the flow transformation eliminate many more phantom peaks in higher dimensions. Comparing the third and fourth column panels, one sees that the output samples of normalizing flow resembles the target distribution, but density distributions are distorted or underestimated in certain regions. This indicates that the learning is imperfect,
 and the resampling is indispensable to fully match the target distribution. 

Figure~\ref{pic:compare} depicts a quantitative performance comparison. The upper panel shows the KL divergence between the sampled and target distributions for uniform sampling, VEGAS sampling, and SG sampling. While uniform and VEGAS sampling rapidly degrade with increasing dimension, SG maintains the lowest KL divergence from $D=2$ up to $D=12$, indicating a faithful reconstruction of the full probability density. The lower panel reports the corresponding uncertainties in the numerical evaluation of the normalization integral. Consistent with the KL results, SG achieves significant reduction in the integral error relative to conventional methods in high dimensions.

\bmhead{Sampling quantum states of atomic nucleus in \texorpdfstring{$D=60,~288,~624$}{D=60, 288, 624} dimensions}
We now turn to a substantially more challenging application for sampling correlated nucleon configurations inside atomic nuclei. This problem is central to nuclear structure calculations and to modeling the initial state of high-energy heavy-ion collisions, where realistic spatial correlations among nucleons play a decisive role~\cite{Li:2025hae,Li:2025bdn}.

For a nucleus with mass number $A$, the many-body configuration space has dimensionality $D=3A$. To examine systems spanning light to heavy nuclei, we consider $^{20}$Ne,  $^{96}$Zr, and $^{208}$Pb, corresponding to $A= 20,\,96,\,208$ and hence $D=60,\,288,\,624$, respectively. Sampling nucleon configurations in such high-dimensional spaces is highly challenging for conventional Monte Carlo approaches, as the underlying probability distributions exhibit strong correlations arising from short-range repulsion, intermediate-range attraction, and the finite-size geometry of the nucleus.

For demonstration, we adopt an effective correlated density of the form
\begin{equation}
f_{A}(\boldsymbol{r}_1,...,\boldsymbol{r}_A) 
=\frac{1}{\mathcal{N}}
\prod_{i=1}^{A}\rho_{WS}(\boldsymbol{r}_i)
\prod_{i<j}g (\boldsymbol{r}_i,\boldsymbol{r}_j),\label{eq:nucleus}
\end{equation}
where the one-body Woods--Saxon (WS) distribution is given by
\begin{equation}
\rho_{WS}(r)
=\rho_0\frac{1+w r^2/R^2}{1+\exp[(r-R)/a]},\label{eq:ws}
\end{equation}
where $R(\theta,\phi)=R_0(1+\Sigma_{lm}  \beta_l Y_{lm}(\theta,\phi))$, and $Y_{lm}(\theta,\phi)$ denotes the spherical harmonic function. For $^{20}$Ne, $^{96}$Zr and $^{208}$Pb, we take WS distribution parameters from Refs.~\cite{Wang:2024ulq,Jia:2022qgl,Rybczynski:2013yba}, as summarized in Table~\ref{TabI}. The average density $\rho_0$ can be absorbed in the normalization factor which does not affect the sampling. The two-body correlation function is parameterized as~\cite{Cruz-Torres:2017sjy}
\begin{equation}
g(\Delta r)
=
1-e^{-\alpha(\Delta r)^2}
\left(
\gamma+\sum_{i=1}^{3}\zeta_i(\Delta r)^{i+1}
\right),~\label{eq:corr}
\end{equation}
with $\Delta r=|\boldsymbol{r}_j-\boldsymbol{r}_k|$ denoting the relative distance between two nucleons. The parameters $\alpha$, $\zeta_i$, and $\gamma$ control the range and strength of correlations. Typical values of parameters in Eq.~(\ref{eq:corr}) are $\alpha =1.26$, $\zeta_1=2.10$, $\zeta_2=-6.13$,~$\zeta_3=2.93$, ~$\gamma=0.992$~\cite{Alvioli:2009ab,Pudliner:1997ck}.
  
\begin{figure}[!t]
\centering 
\includegraphics[width=\linewidth]{ 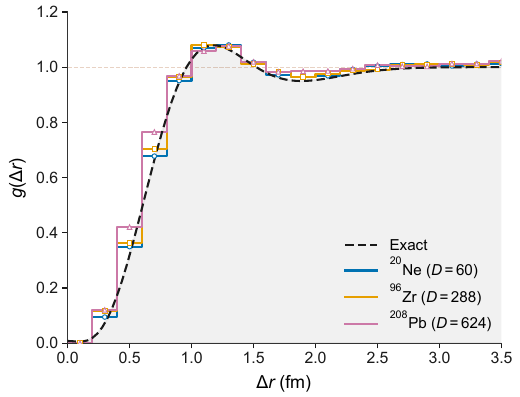} 
\caption{
\textbf{Two-nucleon correlation functions in nuclei with the Schr\"odinger Generator.}
Radial two-body correlation functions $g(\Delta r)$ for nucleus $^{20}$Ne, $^{96}$Zr, and $^{208}$Pb as a function of relative distance $\Delta r$. Dashed lines denote the analytical input correlations, while solid lines represent results extracted from the generated samples in $D=60,\,288,$ and $624$ dimensions.  
}
\label{pic:nucleus} 
\end{figure}

\begin{table}[t]
\centering
\caption{Woods--Saxon distribution parameters taken from Refs~\cite{Wang:2024ulq,Jia:2022qgl,Rybczynski:2013yba}.}
\label{TabI}
\begin{tabular}{l c c c c c}
\toprule
Nucleus & $R_0$ (fm) & $w$ & $a$ (fm) & $\beta_2$ & $\beta_3$ \\
\midrule
$^{20}$Ne   & 2.791  & $-0.168$ & 0.698  & 0.666 & 0.250 \\
$^{96}$Zr   & 5.02   & --       & 0.52   & 0.06  & 0.20\\
$^{208}$Pb  & 6.49   & --       & 0.54   & 0.00  & 0.00 \\
\bottomrule
\end{tabular}
\end{table}

Figure~\ref{pic:nucleus} presents the two-nucleon correlation functions, $g(\Delta r)$, extracted from SG-generated samples for $^{20}$Ne, $^{96}$Zr, and $^{208}$Pb. The dashed lines denote the analytical input correlations, while the solid lines represent results from samples generated in $D=60,\,288,$ and $624$ dimensions, respectively. Even for the heaviest system, $A=208$, the sampled correlations quantitatively reproduce the analytical input across the entire range of relative distances—accurately capturing both the pronounced short-range suppression and the intermediate-distance enhancement.  

To enhance performance for $A=208$, we implement a sequential SG approach by applying two successive operations that each learn a simplified target distribution, $f_{A=104}$. Their product, $f^{(1)}_{104}f^{(2)}_{104}$, serves as the proposal prior for the normalizing flows and resampling, with the importance distribution defined by the ratio $f_{208}/(f^{(1)}_{104}f^{(2)}_{104})$. This hierarchical refinement significantly reduces weight variance and ensures high-fidelity sampling for heavy nucleus.

The above results demonstrate that the Schr\"odinger Generator remains scalable and expressive in configuration spaces exceeding 600 dimensions. This enables simulations across the whole nuclear chart, which would benefit the study of nuclear structure in high-energy nucleus collisions~\cite{Li:2025bdn}.

\bmheadA{Discussion}  
We have developed the Schr\"{o}dinger Generator, a hybrid framework that unifies adaptive Monte Carlo variance reduction with normalizing-flow–based generative modeling through explicit optimization of coordinate transformations. The optimization is achieved by decomposing the total Jacobian into two complementary parts with the first adaptive map that learns the marginal distribution by minimizing variance along individual dimensions and a second flow-based transformation that gradually captures non-trivial correlations across different dimensions.

For benchmarks on fermionic many-body densities, we demonstrate that the Schr\"{o}dinger Generator accurately reconstructs multimodal structures and nodal surfaces while reducing integration uncertainty by more than an order of magnitude relative to conventional Monte Carlo methods. 
In particular, the method effectively removes the phantom peaks commonly encountered in traditional adaptive importance sampling and consistently maps them onto the true modes of the target distribution, without suffering from the mode-collapse problem that often limits standard normalizing flows. Furthermore, we demonstrate stable and scalable performance for nuclear many-body systems up to $D=624$ dimensions, where the Schr\"{o}dinger Generator reproduces short-range correlations and global density profiles, enabling event-by-event simulations of strongly correlated nuclear systems.

Several directions for future development can be envisioned. The framework could be combined with stratified sampling methods~\cite{Lepage:2020tgj} to further reduce variance, especially in regions of phase space with sharp features. Extensions incorporating quantum adaptive importance sampling~\cite{Pyretzidis:2025stx} may provide additional advantages for problems with intrinsically quantum correlations. Besides, replacing or augmenting the current flow component with continuous normalizing flows~\cite{Chen:2018wjc,jt6y-h375} trained with Flow Matching method~\cite{lipman2022flow,Albergo:2022iol,Liu:2022yjp,Albergo:2025ihk} could improve the handling of complex geometries and topologies by enabling smoother and more flexible transformations. 
All together, the framework of Schr\"{o}dinger Generator offers a general strategy for high-dimensional stochastic integration and unbiased sampling, with applications in Bayesian inference, machine learning, statistical mechanics, and quantum many-body problems involving highly-structured probability distributions.

\bmheadA{Methods}  
The Schr\"odinger Generator comprises three tightly coupled components:
(i) an adaptive coordinate transformation that reshapes the integration domain to minimize estimator variance in each dimension separately;
(ii) a normalizing flow that learns high-dimensional correlations across different dimensions and enables expressive sampling; and
(iii) a resampling step that guarantees unbiased sampling even when the learned transformation is not ideal.
 Detailed realization of this algorithm is described below.

\bmhead{1. Adaptive mapping and iterative refinement of grids} 
To determine the adaptive Jacobian component $J_A$, we follow the conventional \textsc{VEGAS} scheme~\cite{Lepage:1977sw} and set $p_0(z)=U[0,1]$.
Starting from one-dimension integral, the domain $[0,1]$ in $z$-space is divided into $N_A$ equal intervals
$\{[z_i,z_i+\Delta z_i]\}$, which are mapped onto intervals $\{[y_i,y_i+\Delta y_i]\}$ in $y$-space and then mapped identically onto intervals in $x-$space.
The Jacobian is then given by
\begin{equation}
J_A(z) = N_A\,\Delta x_{i(z)} 
\end{equation}
where ${i(z)}$ denotes the integer part of $N_A \times z$.
The variance can be written as
\begin{equation}
\sigma_I^2 = \frac{1}{M}\left [\sum_i J_i
\int_{x_i}^{x_i+\Delta x_i}\!\!\mathrm{d}x\, f^2(x)
- I^2 \right ].
\end{equation}
Treating the $J_i$ as independent variables subject to the constraint
$\sum_i \Delta x_i/J_i = 1$, the variance is minimized when~\cite{Lepage:1977sw}
\begin{equation}
\frac{J_i^2}{\Delta x_i}
\int_{x_i}^{x_i+\Delta x_i}\!\!\mathrm{d}x\, f^2(x)
=\text{const}.
\label{eq:discrete_optimal}
\end{equation}

This condition is realized iteratively.
Starting from a uniform grid, Monte Carlo samples are used to estimate
\begin{equation}
d_i = \langle J^2 f^2\rangle_{\Delta x_i},
\end{equation}
where the average is taken for Monte-Carlo samples inside the $i-$th interval [$x_i,x_i+\Delta x_i$].
The grid is then updated so that each interval carries an equal fraction of the accumulated $d_i$. To ensure stability, the estimates are smoothed and compressed before each update, suppressing statistical noise and preventing overreaction to rare fluctuations. This adaptive procedure becomes exact only when the integrand can be factorized in each dimension, which, however, fails in the presence of correlations.

After fixing $J_A$, samples in $z-$space are generated uniformly in $[0,1]^D$, and they are then mapped to $y-$space according to the coordinate transformation. To capture non-factorizable correlations, the Schr\"odinger Generator applies a normalizing flow transformation from $y-$space to $x-$space.

\bmhead{2. Normalizing flow for correlation learning}
The normalizing flow transformation is given by a series of invertible bijections,
\begin{equation}
\mathbf{x} \equiv \mathbf{x}^{(K+1)} = c_K \circ \cdots \circ c_1(\mathbf{x}^{(1)}=\mathbf{y}),
\end{equation}
with total Jacobian given by
\begin{equation}
J_F
=
\prod_{k=1}^K
\left|
\det\frac{\partial c_k}{\partial \mathbf{x}^{(k-1)}}
\right|^{-1}.
\end{equation}

We employ coupling-layer–based flows~\cite{Dinh:2014mzt,Muller:2018pvg,Durkan:2019nsq} in which the variables are partitioned into
$\mathbf{x}^{(1)}=(\mathbf{x}^{(1)}_{A},\mathbf{x}^{(1)}_{B})$.
We use two-step (bidirectional) coupling:
\begin{eqnarray} 
\tilde{\mathbf{x}}^{(1)}_{A} &=& T_{\rm RQS}(\mathbf{x}^{(1)}_{A}, \theta^{(1)}_A), \notag \\
\tilde{\mathbf{x}}^{(1)}_{B} &=& T_{\rm RQS}(\mathbf{x}^{(1)}_{B}, \theta^{(1)}_B),\label{eq:rqs}
\end{eqnarray} 
where the parameters ($\theta=\{  w,h,\delta \}$) are given by the output of full-connected neural network, i.e., $\theta^{(1)}_A = FCNN(\mathbf{x}^{(1)}_{B})$ and $\theta^{(1)}_B = FCNN(\tilde{\mathbf{x}}^{(1)}_{A})$. The output is combined as $\mathbf{x}^{(2)} \equiv c_1(\mathbf{x}^{(1)})=(\tilde{\mathbf{x}}^{(1)}_{A}, \tilde{\mathbf{x}}^{(1)}_{B})$.
This sequential structure preserves a triangular Jacobian and enables efficient evaluation of the log-determinant.

In Eq.~(\ref{eq:rqs}), $T_{\rm RQS}$ denotes a monotonic, invertible mapping $T_{\rm RQS}:\mathbb{R}\to\mathbb{R}$ based on a rational quadratic spline defined on a compact interval $[-B,B]$~\cite{Durkan:2019nsq} . For a scalar transformation $t = T_{\rm RQS}(s,\theta)$, the domain is partitioned into $J$ bins with knots $\{s_j\}_{j=0}^J$ and $\{t_j\}_{j=0}^J$, satisfying $s_0=t_0=-B$ and $s_J=t_J=B$. The bin widths $w_j = s_j - s_{j-1}$ and heights $h_j = t_j - t_{j-1}$ obey
$\sum_{j=1}^J w_j = 2B, \qquad \sum_{j=1}^J h_j = 2B$, and define slopes $g_j = h_j / w_j$. In addition, strictly positive derivatives $\{\delta_j\}_{j=0}^J$ are specified at the knots.
 
For $s \in [s_j, s_{j+1}]$, introducing the normalized coordinate $\xi = {(s - s_j)}/{w_j} \in [0,1]$,
the transformation is given by
\begin{eqnarray}  
T_{\rm RQS}(s)
&=&\sum_{j=1}^J
(t_j
+
\frac{h_j \left[ g_j \xi^2 + \delta_j \, \xi(1-\xi) \right]}
{g_j + (\delta_{j+1} + \delta_j - 2g_j)\, \xi(1-\xi)})\notag \\
&\times& (H(s-s_j)-H(s-s_{j+1})),
\label{eq:RQS}
\end{eqnarray}
where $H$ denotes the step function. 
The corresponding Jacobian in the interval $[s_j,s_{j+1}]$ is obtained as
\begin{equation}
\frac{dT_{\rm RQS}}{ds}
=
\frac{
g_j^2 \left[
\delta_{j+1} \xi^2
+
2 g_j \xi(1-\xi)
+
\delta_j (1-\xi)^2
\right]
}{
\left[
g_j + (\delta_{j+1} + \delta_j - 2g_j)\, \xi(1-\xi)
\right]^2
}.
\label{eq:RQS_derivative}
\end{equation} 
This construction ensures continuity of both the function and its first derivative across bin boundaries, while strict positivity of $\{w_j, h_j, \delta_j\}$ guarantees global monotonicity and hence invertibility. The inverse mapping $T_{\rm RQS}^{-1}$ is obtained by solving Eq.~(\ref{eq:RQS}) for $\xi$, which reduces to a quadratic equation, followed by $s = s_j + \xi w_j$.

Outside the interval $[-B,B]$, the transformation is extended linearly to preserve invertibility. In practice, the parameters $\{w_j, h_j, \delta_j\}$ are generated by neural networks from unconstrained outputs via softmax (for $w_j, h_j$) and softplus (for $\delta_j$), ensuring the required positivity and normalization constraints. The spline coupling thus provides an expressive yet analytically tractable nonlinear transformation. In each layer, the activation normalization and one-by-one convolution are adopted to stabilize the optimization.

\begin{algorithm}[t]
\caption{Schr\"odinger Generator}
\label{alg:SG}
~\textbf{Input:} Target density $f(\mathbf{x})$ and sample size $M$\\
\textbf{Output:} Monte Carlo estimate of integral and samples from $f$.

Initialize adaptive grid for $\mathcal{T}_A$ and flow parameters for $\mathcal{T}_F$\;

(i) \For{each training iteration}{
    Sample $\mathbf{z}_i \sim U[0,1]$\;
    Transform $\mathbf{y}_i = \mathcal{T}_A(\mathbf{z}_i)$ \;$\mathbf{x}_i$=$\mathbf{y}_i$\;
    Compute weights 
    $w_{A,i} = f(\mathbf{x}_i) J_A(\mathbf{z}_i)$\;
    Adapt grid bin widths using accumulated $d_i=w_{A,i}^2$ statistics to minimize variance loss~\cite{Lepage:2020tgj}\; }

(ii) \For{each training iteration}{
    Sample $\mathbf{z}_i \sim U[0,1]$\;
    Transform $\mathbf{y}_i = \mathcal{T}_A(\mathbf{z}_i)$, 
    $\mathbf{x}_i = \mathcal{T}_F(\mathbf{y}_i)$\;
    Compute weights 
    $w_{SG,i} = f(\mathbf{x}_i) J_F(\mathbf{y}_i)J_A(\mathbf{z}_i)$\; 
    Update flow parameters by minimizing $\mathrm{KL}$ loss\;
    
}

(iii) Estimate integral 
$I_{\rm SG} = \frac{1}{M}\sum_i w_{SG,i}$ and uncertainty $\sigma_{I_{\rm SG}} = \sqrt{\frac{1}{M-1} \left(\frac{1}{M}\sum_i w_{SG,i}^2-I_{\rm SG}^2\right) }$, and resample $\mathbf{x}$ with probability $\propto w_{SG}$ to obtain exact samples from $f$\;

\Return $I_{\rm SG}$, $\sigma_{I_{\rm SG}}$, and samples $\{ \mathbf{x}_{i} \}$\;
\end{algorithm}

Flow parameters are optimized by minimizing the Kullback–Leibler divergence~\cite{Kullback:1951zyt} between
the transformed distribution and the target density,
\begin{eqnarray}
\mathcal{L}_{\mathrm{SG}}
=
-\mathbb{E}
 [\log f(\mathbf{x}\circ\mathbf{y}(\mathbf{z}))+
\log J_F +\log J_A ].  
\end{eqnarray}

\bmhead{3. Integration and Resampling}
The full transformation Jacobian is $J_{SG} = J_A J_F$.
To guarantee unbiased sampling when $J_{SG}$ is not optimal,
latent samples are assigned importance weights
$w_{SG}(\mathbf{x})=f(\mathbf{x}(\mathbf{z}))\,J_{SG}(\mathbf{z})$.
The Monte Carlo estimator of integral and its uncertainty are given by
\begin{eqnarray}
I_{\rm SG}&=&\frac{1}{M}\sum_i w_{SG,i}, \notag \\
\sigma_{I_{\rm SG}} &=& \sqrt{\frac{1}{M-1} \left(\frac{1}{M}\sum_i w_{SG,i}^2-I_{\rm SG}^2\right) },
\end{eqnarray}
and final samples are obtained by resampling with probability proportional to $w_{SG}$. The complete procedure is summarized in Algorithm~\ref{alg:SG}. 

\bmhead{Acknowledgments} The author K. J. Sun thanks helpful discussions with Peng-Sheng Wen, Bo Zhou, Long-Gang Pang, and Kai Zhou.
This work was supported in part by the National Key Research and Development Project of China under Grant No. 2024YFA1612500 and No. 2022YFA1604900; the National Natural Science Foundation of China under contract No. 12422509, No. 12375121, No. 12547102, No. 12325507, and No. 12147101. The computations in this research were performed using the CFFF platform of Fudan University.  

\bmhead{Data availability}
All the data supporting the findings in this work are available within the manuscript and any additional data are available from the corresponding authors upon reasonable request.

\bmhead{Code availability}
The codes supporting this study are openly available in a GitHub repository at \href{https://github.com/kjsunHIC/SchrodingerGenerator}{github.com/kjsunHIC}.

\bmhead{Author contributions}
L. J. Jiang, F. Ma, and P. Li performed the numerical simulations, prepared the figures, and contributed equally to this work. K. J. Sun supervised the project. All authors contributed to the discussions and to the preparation of the manuscript.

\bmhead{Competing interests}
The authors declare no competing interests.  




\end{document}